\documentstyle[12pt,epsf]{article}

\newcommand{\U}{{\rm U}}
\newcommand{\vect}{{\rm Vect}}
\newcommand{\Ref}[1]{(\ref{#1})}
\newcommand{\YM}{{\cal Y\!M}}

\newcommand{\R}{\real}

\renewcommand{\thefootnote}{\fnsymbol{footnote}}
\newcommand{\newsection}{\setcounter{equation}{0}\section}

\def\appendix#1{\addtocounter{section}{1}\setcounter{equation}{0}
\renewcommand{\thesection}{\Alph{section}}
\section*{Appendix \thesection\protect\indent \parbox[t]{11.715cm} {#1}}
\addcontentsline{toc}{section}{Appendix \thesection\ \ \ #1} }

\newcommand{\complex}{{\bb C}} 
\newcommand{\real}{{\bb R}} 
\newcommand{\reals}{{\bbs R}} 

\newif\ifold             \oldtrue

\font\mybb=msbm10 at 12pt
\def\bb#1{\hbox{\mybb#1}}
\font\mybbs=msbm10 at 9pt
\def\bbs#1{\hbox{\mybbs#1}}

\def\nn{\nonumber}

\def\be{\begin{equation}}
\def\ee{\end{equation}}
\def\bea{\begin{eqnarray}}
\def\eea{\end{eqnarray}}
\def\bd{\begin{displaymath}}
\def\ed{\end{displaymath}}

\newcommand{\beq}{\begin{eqnarray}}
\newcommand{\eeq}{\end{eqnarray}}

\newcommand{\ka}{ e}

\begin{document}
\begin{titlepage}
\begin{flushright}

\baselineskip=12pt

HWM--01--15\\
EMPG--01--05\\
hep--th/0105094\\
\hfill{ }\\
May 2001
\end{flushright}

\begin{center}

\baselineskip=24pt

\vspace{1cm}

{\Large\bf Teleparallel Gravity and Dimensional Reductions \\ of Noncommutative
Gauge Theory}

\baselineskip=14pt

\vspace{1cm}

{\bf Edwin Langmann}
\\[4mm]
{\it Department of Theoretical Physics, Royal Institute of Technology\\ S-100
44 Stockholm,
Sweden}\\{\tt langmann@theophys.kth.se}
\\[10mm]

{\bf Richard J. Szabo}
\\[4mm]
{\it Department of Mathematics, Heriot-Watt University\\ Riccarton, Edinburgh
EH14 4AS, Scotland}
\\{\tt R.J.Szabo@ma.hw.ac.uk}
\\[30mm]

\end{center}

\begin{abstract}

\baselineskip=12pt

We study dimensional reductions of noncommutative electrodynamics on flat
space which lead to gauge theories of gravitation. For a
general class of such reductions, we show that the noncommutative gauge
fields naturally yield a Weitzenb\"ock geometry on spacetime and that the
induced diffeomorphism invariant field theory can be made equivalent
to a teleparallel formulation of gravity which macroscopically describes
general relativity. The Planck length is determined in this setting by the
Yang-Mills coupling constant and the noncommutativity scale. The effective
field theory can also contain higher-curvature and non-local terms which are
characteristic of string theory. Some applications to D-brane dynamics and
generalizations to include the coupling of ordinary Yang-Mills theory to
gravity are also described.

\end{abstract}

\end{titlepage}
\setcounter{page}{2}

\newpage

\renewcommand{\thefootnote}{\arabic{footnote}} \setcounter{footnote}{0}

\newsection{Introduction}

Yang-Mills theory on a flat noncommutative space arises as special decoupling
limits of string theory~\cite{NCYMstring,SW} and M-theory~\cite{SW,CDS}. In
string theory it represents the low-energy effective field theory induced on
D-branes in the presence of a constant background supergravity $B$-field. The
inherent non-locality of the interactions in this field theory lead to many
exotic effects that do not arise in ordinary quantum field theory, and which
can be attributed to ``stringy'' properties of the model. It is thereby
believed that, as field theories, these models can provide an effective
description of many of the non-local effects in string theory, but within a
much simpler setting. As string theory is a candidate for a unified quantum
theory of the fundamental interactions, and in particular of gravitation, it is
natural to seek ways to realize this unification in the context of
noncommutative gauge theories. Gravity has been previously discussed using the
framework of noncommutative geometry in~\cite{CFF}, while the unification of
Einstein gravity and Yang-Mills theory is obtained in~\cite{ChamConnes} from a
spectral action defined on an almost-commutative geometry. In this paper we
will describe a particular way that gravitation can be seen to arise in
noncommutative Yang-Mills theory on {\it flat} space.

There are several hints that gravitation is naturally contained in the gauge
invariant dynamics of noncommutative Yang-Mills theory. In~\cite{IIKKString}
the strong coupling supergravity dual of maximally supersymmetric
noncommutative Yang-Mills theory in four dimensions~\cite{dualgrav} was studied
and it was shown that the effective supergravity Hamiltonian has a unique zero
energy bound state which can be identified with a massless scalar field in four
dimensions. The ten dimensional supergravity interaction is then of the form of
a four dimensional graviton exchange interaction and one may therefore identify
the Newtonian gravitational potential in noncommutative gauge theory. The
Planck length is determined in this setting by the scale of noncommutativity.
Furthermore, noncommutative Yang-Mills theory at one-loop level gives rise to
long-range forces which can be interpreted as gravitational interactions in
superstring theory~\cite{AIIKKT,LongRange}.

At a more fundamental level, general covariance emerges in certain ways from
the extended symmetry group that noncommutative gauge theories possess. It can
be seen to emerge from the low-energy limit of a closed string vertex operator
algebra as a consequence of the holomorphic and anti-holomorphic mixing between
closed string states~\cite{LLS}. The diffeomorphism group of the target space
acts on the vertex operator algebra by inner automorphisms and thereby
determines a gauge symmetry of the induced noncommutative gauge theory.
Furthermore, noncommutative Yang-Mills theory can be nonperturbatively
regularized and studied by means of twisted large $N$ reduced
models~\cite{AIIKKT},\cite{Li}--\cite{AMNS2}. This correspondence identifies
the noncommutative gauge group as a certain large $N$ limit~\cite{llsLargeN} of
the unitary Lie group ${\rm U}(N)$ which is equivalent to the symplectomorphism
group of flat space~\cite{FZ}. The noncommutative gauge group can thereby be
described as a certain deformation of the symplectomorphism group (or
equivalently ${\rm U}(\infty)$)~\cite{LSZPrep} and the noncommutative gauge
theory can be regarded as ordinary Yang-Mills theory with this extended,
infinite dimensional gauge symmetry group~\cite{MMS1loop}. Other attempts at
interpreting diffeomorphism invariance in noncommutative Yang-Mills theory
using reduced models of gauge theory can be found in~\cite{CS,AIKO}.

These features are all consequences of the fact that noncommutative gauge
transformations mix internal and spacetime symmetries, and are thereby very
different from ordinary gauge symmetries. In the case of noncommutative
Yang-Mills theory on flat infinite space, a global translation of the spacetime
coordinates can be realized as a local gauge
transformation~\cite{AMNS2,GN1,Transl}, up to a global symmetry of the field
theory. The main consequence of this property is that all gauge invariant
operators are non-local in the sense that their translational invariance
requires them to be averaged over spacetime. Such constructions are reminescent
of general relativity. In addition, the noncommutative gauge symmetry allows
for extended gauge invariant operators~\cite{Transl} which are constructed from
open Wilson line observables~\cite{AMNS1,AMNS2,IIKKWilson}. These observables
exhibit many of the ``stringy'' features of noncommutative gauge
theory~\cite{Transl,Rozvan}. They can also be used to construct the appropriate
gauge invariant operators that couple noncommutative gauge fields on a D-brane
to massless closed string modes in flat space~\cite{WilsonGrav}, and thereby
yield explicit expressions for gauge theory operators dual to bulk supergravity
fields in this case.

The fact that the group of global translations is contained in the group of
noncommutative gauge transformations is thereby naturally linked with the
suggestion that noncommutative Yang-Mills theory may contain gravity. The idea
of representing general relativity as a gauge theory of some kind is of course
an old idea~\cite{Kibble}--\cite{Hehlrev} (See~\cite{Hehl} for more recent
reviews of the gauge theory approaches to gravitation). Such models are based
on constructing gauge theories with structure groups given by spacetime
symmetry groups, such as the Poincar\'e group, in such a way that the mixing of
gauge and spacetime symmetries enables the unambiguous identification of gauge
transformations as general coordinate transformations. If noncommutative gauge
theory is to contain gravitation in a gauge invariant dynamical way, then its
gauge group should admit a local translational symmetry corresponding to
general coordinate transformations in flat space. While there are general
arguments which imply that this is not the case~\cite{LSZPrep,AIKO}, it could
be that a particular reduction of noncommutative gauge theory captures the
qualitative manner in which noncommutative gauge transformations realize
general covariance. In this paper we shall discuss one such possibility. We
will show how noncommutative $\U(1)$ Yang-Mills theory on {\it flat} space
$\R^{n}\times\R^{n}$ can generate a theory of gravitation on $\R^n$. The basic
observation underlying the construction is that the algebra of functions on
$\R^{2n}$, with Lie bracket defined in terms of the deformed product of the
noncommutative theory, contains the Lie algebra $\vect(\R^n)$ of vector fields
on $\R^n$.\footnote{\baselineskip=12pt By diffeomorphism invariance in the
following we will mean invariance under the connected diffeomorphism group,
i.e. under the Lie algebra $\vect(\reals^n)$ of infinitesimal diffeomorphisms.
In this paper we will not consider any global aspects of the gauge symmetries.}
We show that it is possible to consistently restrict the noncommutative
Yang-Mills fields so as to obtain a local field theory whose symmetry group
contains diffeomorphism invariance. This construction shows how noncommutative
gauge symmetries give very natural and explicit realizations of the mixing of
spacetime and internal symmetries required in the old gauge models of gravity.
It is also reminescent of a noncommutative version of brane world constructions
which localize ten dimensional supergravity down to four dimensions along some
noncommutative directions~\cite{NCWorld}.

Gauge theories whose structure group is the group of translations of spacetime
lead to teleparallel theories of gravity~\cite{Tele}. These models are built
via an explicit realization of Einstein's principle of absolute parallelism.
They are defined by a non-trivial vierbein field which can be used to construct
a linear connection that carries non-vanishing torsion, but is flat. Such a
connection defines what is known as a Weitzenb\"ock geometry on spacetime. The
vanishing of the curvature of the connection implies that parallel transport in
such a geometry is path-independent, and so the geometry yields an absolute
parallelism. Teleparallelism thereby attributes gravitation to torsion, rather
than to curvature as in general relativity. This class of gravitational
theories is thereby a very natural candidate for the effective noncommutative
field theory of gravitation, which is induced on flat spacetime. We will see in
the following how the gauge fields of the dimensionally reduced noncommutative
Yang-Mills theory naturally map onto a Weitzenb\"ock structure of spacetime. A
teleparallel theory of gravity can also be viewed as the zero curvature
reduction of a Poincar\'e gauge theory~\cite{Kibble,Chogen,Hehlrev} which
induces an Einstein-Cartan spacetime characterized by connections with both
non-vanishing torsion and curvature. The zero torsion limit of an
Einstein-Cartan structure is of course a Riemannian structure and is associated
with ordinary Einstein general relativity. The Weitzenb\"ock geometry is in
this sense complementary to the usual Riemannian geometry. More general gauge
theories of gravitation can be found in~\cite{Hehl,Gengrav}. From the present
point of view then, noncommutative Yang-Mills theory naturally induces
gravitation through a torsioned spacetime, and its full unreduced dynamics may
induce gravity on the entire spacetime through the gauging of some more
complicated spacetime group, as in~\cite{Gengrav}.

A teleparallel gauge theory of gravity describes the dynamical content of
spacetime via a Lagrangian which is quadratic in the torsion tensor
$T_{\mu\nu\lambda}$ of a Cartan connection. The most general such Lagrangian is
given by\footnote{\baselineskip=12pt In this paper an implicit summation over
repeated upper and lower indices is always understood, except when noted
otherwise.}
\beq
{\cal L}_T=\frac1{16\pi\,G_{\rm
N}}\,\Bigl(\tau_1\,T_{\mu\nu\lambda}\,T^{\mu\nu\lambda}
+\tau_2\,T_{\mu\nu\lambda}\,T^{\mu\lambda\nu}+\tau_3\,T_{\mu\nu}^{~~\nu}\,
T^{\mu\lambda}_{~~\lambda}\Bigr) \ ,
\label{calLT}\eeq
where $G_{\rm N}$ is the Newtonian gravitational constant and $\tau_i$,
$i=1,2,3$, are arbitrary parameters. For generic values of $\tau_i$ the field
theory defined by (\ref{calLT}) is diffeomorphism invariant, but it is not
equivalent to Einstein gravity~\cite{Cho}. However, one can demand that the
theory (\ref{calLT}) yield the same results as general relativity in the
linearized weak field approximation. It may be shown that there is a
one-parameter family of Lagrangians of the form (\ref{calLT}), defined by the
parametric equation
\beq
2\tau_1+\tau_2=1 \ , ~~ \tau_3=-1 \ ,
\label{GRpareqns}\eeq
which defines a consistent theory that agrees with all known gravitational
experiments~\cite{Teleexpt}. For such parameter values the Lagrangian
(\ref{calLT}) represents a physically viable gravitational theory which is
empirically indistinguishable from general relativity. For the particular
solution $\tau_1=\frac14,\tau_2=\frac12$ of (\ref{GRpareqns}), the Lagrangian
(\ref{calLT}) coincides, modulo a total divergence, with the Einstein-Hilbert
Lagrangian~\cite{Hehlrev,Hehl,TeleGR}, and the resulting gauge theory is
completely equivalent to Einstein gravity at least for macroscopic, spinless
matter. In addition, the Weitzenb\"ock geometry possesses many salient features
which makes it particularly well-suited for certain analyses. For instance, it
enables a pure tensorial proof of the positivity of the energy in general
relativity~\cite{Nester}, it yields a natural introduction of Ashtekar
variables~\cite{Mielke}, and it is the naturally setting in which to study
torsion~\cite{Okubo}, notably at the quantum level, in systems whereby torsion
is naturally induced, such as in the gravitational coupling to spinor fields.

In this paper we will show that the dimensional reduction of noncommutative
gauge theory that we consider contains the Lagrangian (\ref{calLT}) for the
particular family of teleparallel theories of gravity defined by
$\tau_2=\tau_3$. Thus for the solution $\tau_1=-\tau_2=1$ of (\ref{GRpareqns}),
it contains a macroscopic description of general relativity. Whether or not
these latter constants really arise is a dynamical issue that must be treated
by regarding the dimensionally reduced noncommutative field theory as an
effective theory, induced from string theory for example. We shall not address
this problem in the present analysis, except to present a group theoretical
argument for the naturality of this choice of parameters. From this result we
will determine the gravitational constant in terms of the gauge coupling
constant and the noncommutativity scale. We will also find a host of other
possible terms in the total Lagrangian which we will attribute to
higher-curvature and non-local couplings that are characteristic of string
theory. Indeed, the particular theory induced by the standard, flat space
noncommutative Yang-Mills theory on a D-brane is a special case of the more
general construction. In that case we will find quite naturally that the
gravitational theory can be
invariant only under the volume-preserving coordinate transformations of
spacetime, a fact anticipated from string theoretical considerations. We will
also describe how the present construction can be generalized to include the
coupling of gravity to ordinary gauge fields. These results all show that, at
the level of the full unreduced Yang-Mills theory, noncommutative gauge
symmetry naturally contains gravitation and also all other possible commutative
gauge theories, at least at the somewhat simplified level of dimensional
reduction and the principle of absolute parallelism. The constructions shed
some light on how the full gauge invariant dynamics of noncommutative
Yang-Mills theory incorporates gravitation. At a more pragmatic level,
noncommutative Yang-Mills theories give a very natural and systematic way of
inducing gauge models of gravity in which the mixing between spacetime and
internal degrees of freedom is contained in the gauge invariant dynamics from
the onset.

The structure of the remainder of this paper is as follows. In section~2 we
describe the general model of noncommutative $\U(1)$ Yang-Mills theory and its
gauge invariant dimensional reductions. In section~3 we describe a particular
family of dimensional reductions and compute the induced actions. In section~4
we construct a Weitzenb\"ock structure on spacetime from the dimensionally
reduced gauge fields, and in section~5 we relate the leading low-energy
dynamics of the induced Lagrangian to a teleparallel theory of gravity. In
section~6 we specialize the construction to ``naive'' dimensional reductions
and describe the natural relationships to D-branes and volume-preserving
diffeomorphisms. In section~7 we describe the dynamics induced by certain
auxilliary fields which are required to complete the space of noncommutative
gauge fields under the reductions. We show that they effectively lead to
non-local effects, which are thereby attributed to stringy properties of the
induced gravitational model. In section~8 we describe how to generalize the
construction to incorporate ordinary gauge fields coupled to the induced
gravity
theory, and in section~9 we conclude with some possible extensions and further
analyses of the model presented in this paper. An appendix at the end of the
paper contains various identities which are used to derive quantities in the
main text.

\newsection{Generalized Noncommutative Electrodynamics}

Consider noncommutative $\U(1)$ Yang-Mills theory on flat Euclidean
space $\real^{2n}$, whose local coordinates are denoted
$\xi=(\xi^A)_{A=1}^{2n}$. The star-product on the algebra
$C^\infty(\real^{2n})$ of smooth functions
$f,g:\real^{2n}\to\complex$ is defined by\footnote{\baselineskip=12pt Later,
when we come to the construction of action functionals, we will need to
restrict the space $C^\infty(\reals^{2n})$ to its subalgebra consisting of
functions which decay sufficiently fast at infinity. Such restrictions can be
imposed straightforwardly and so we will not always spell this out explicitly.}
\beq
f\star g(\xi)=f(\xi)~(\exp\triangle)~g(\xi) \ ,
\label{starproddef}\eeq
where $\triangle$ is the skew-adjoint bi-differential operator
\beq
\triangle=\frac12\,\sum_{A<B}\Theta^{AB}\left(\overleftarrow{\partial}_A
\overrightarrow{\partial}_B-\overleftarrow{\partial}_B
\overrightarrow{\partial}_A\right) \ ,
\label{Deltadef}\eeq
$\partial_A=\partial/\partial\xi^A$, and $\Theta^{AB}=-\Theta^{BA}$ are
real-valued deformation parameters of mass dimension $-2$. The star-product is
defined such that (\ref{starproddef}) is real-valued if the functions $f$ and
$g$ are. It is associative, noncommutative, and it satisfies the usual Leibniz
rule with respect to ordinary differentiation. This implies, in the usual way,
that the
star commutator
\beq
[f,g]_\star= f\star g- g\star f
\label{starcomm_def}
\eeq
defines a Lie algebra structure on $C^\infty(\real^{2n})$. In particular, it
satisfies the Leibniz rule and the Jacobi identity. The star-product
of a function with itself can be represented as
\beq
f\star f(\xi)=f(\xi)~(\cosh\triangle)~f(\xi) \ ,
\label{starself}
\eeq
while the star commutator of two functions is given by
\beq
[f,g]_\star(\xi)=f(\xi)~(2\sinh\triangle)~g(\xi) \ .
\label{starcomm}\eeq

Let us consider the space $\YM$ of $\U(1)$ gauge fields on $\real^{2n}$,
\beq
{\cal A}={\cal A}_A~d\xi^A \ ,
\label{gaugefield}\eeq
where ${\cal A}_A\in C^\infty(\real^{2n})$ is a real-valued function.
Let ${\bf g}\subset C^\infty(\real^{2n})$ be a linear
subalgebra of functions, closed under star commutation, which parametrize the
infinitesimal, local star-gauge transformations defined by
\beq
\delta_\alpha{\cal A}_A=\partial_A\alpha+
e~{\rm ad}_\alpha({\cal A}_A)
\ , ~~\alpha\in{\bf g} \ ,
\label{stargaugetransf}\eeq
where
\beq
{\rm ad}_\alpha(f)=[\alpha,f]_\star \ , ~~ f\in C^\infty(\real^{2n}) \ ,
\label{adalphaf}\eeq
denotes the adjoint action of the Lie algebra $\bf g$ on
$C^\infty(\real^{2n})$. We will require that the space $\YM$ is invariant under
these transformations. The gauge coupling constant $\ka$ in
\Ref{stargaugetransf} will be related later on to the gravitational coupling
constant. The gauge transformation (\ref{stargaugetransf}) is defined such that
the skew-adjoint covariant derivative
\beq
{\cal D}_A=\partial_A+ \ka~{\rm ad}_{{\cal A}_A}
\label{covderiv}
\eeq
has the simple transformation property $\delta_\alpha{\cal D}_A=e\,{\cal
D}_A\alpha$. This and the properties of the star product imply that the linear
map $\alpha\mapsto\delta_\alpha$ is a representation of the Lie algebra ${\bf
g}$,
\beq
[\delta_\alpha,\delta_\beta]=\delta_{ [\alpha,\beta]_\star}~~~~\forall
\alpha,\beta\in{\bf g} \ .
\label{deltastarcomm}\eeq
It also implies as usual that the noncommutative field strength tensor, defined
by
\beq
{\cal F}_{AB}= \frac{1}{ e}\,
\Bigl({\cal D}\wedge{\cal D}\Bigr)_{AB}=\partial_A{\cal A}_B-
\partial_B{\cal A}_A+  e\,\Bigl[{\cal A}_A\,,\,{\cal A}_B\Bigr]_\star \ ,
\label{fieldstrength}\eeq
transforms homogeneously under star-gauge transformations,
\beq
\delta_\alpha{\cal F}_{AB}= {\rm ad}_\alpha({\cal F}_{AB}) \ .
\label{gaugetransfcalF}\eeq

Let $G_{AB}$ be a flat metric on $\R^{2n}$. Then, since the star commutator
(\ref{starcomm}) is a total derivative, the standard action for noncommutative
Yang-Mills theory, defined by
\beq
I_{\rm NCYM}=\frac12~\int\limits_{\reals^{2n}}
d^{2n}\xi~\sqrt{\det G}~G^{AA'}G^{BB'}\,{\cal F}_{AB} \star {\cal F}_{A'B'}
(\xi)\ ,
\label{NCYMstandard}\eeq
is trivially gauge invariant. In the following we will consider reductions of
noncommutative gauge theory on $\real^{2n}$ by imposing certain constraints on
the spaces $\YM$ and ${\bf g}$ and using a generalized action of the form
\beq
I_{\rm NCYM}^{W}=\frac12~\int\limits_{\reals^{2n}}d^{2n}\xi~W^{AA'BB'}(\xi)\,
{\cal F}_{AB}\star {\cal F}_{A'B'}(\xi) \ .
\label{NCYMactiongen}
\eeq
Here $W^{AA'BB'}(\xi)$ are tensor weight functions of rank four with the
symmetries
\beq
W^{AA'BB'}=W^{A'ABB'}=W^{AA'B'B}=W^{BB'AA'} \ .
\label{Wsyms}\eeq
We will require that they transform under star gauge transformations
(\ref{stargaugetransf}) such that the action (\ref{NCYMactiongen}) is gauge
invariant. A sufficient condition for this is
\beq
\int\limits_{\reals^{2n}}d^{2n}\xi~
\Bigl(\delta_\alpha W^{AA'BB'}(\xi)\,f(\xi)+W^{AA'BB'}(\xi)~
{\rm ad}_\alpha(f)(\xi)\Bigr)= 0
\label{condition}\eeq
for all functions $\alpha\in{\bf g}$, $f\in C^\infty(\real^{2n})$ and for each
set of indices $A,A',B,B'$. The functions $W^{AA'BB'}$ generically break the
global Lorentz symmetry which is possessed by the conventional action
(\ref{NCYMstandard}). They are introduced in order to properly maintain gauge
invariance and Lorentz invariance in the ensuing dimensional reductions. The
basic point is that the field strength tensor ${\cal F}_{AB}$ which appears in
(\ref{NCYMstandard}) corresponds to an irreducible representation, i.e. the
rank two antisymmetric representation $\Lambda^2(2n)$, of the Lorentz group of
$\real^{2n}$. This will not be so after dimensional reduction, and the tensor
densities will essentially enforce the decomposition of the reduced field
strengths into irreducible representations of the reduced Lorentz group which
may then be combined into the required singlets. Note that the key to this is
that, because of (\ref{gaugetransfcalF}) and (\ref{condition}), each term in
(\ref{NCYMactiongen}) is individually gauge invariant. The condition
(\ref{condition}) will be used later on to determine an explicit form for the
tensor density $W^{AA'BB'}$ in terms of the noncommutative gauge fields.

We will also consider the minimal coupling of the noncommutative gauge theory
(\ref{NCYMactiongen}) to scalar matter fields. The standard method can be
generalized in a straightforward manner. We assume that the scalar bosons are
described by
real-valued functions $\Phi\in C^\infty(\R^{2n})$ which transform under the
infinitesimal adjoint action of the star gauge symmetry group,
\beq
\delta_\alpha\Phi={\rm ad}_\alpha(\Phi) \ , ~~ \alpha\in{\bf g} \ .
\label{Phigaugetransf}\eeq
Then, by the usual arguments, the action
\beq
I_{\rm B} = \frac12~\int\limits_{\reals^{2n}}d^{2n}\xi~
W(\xi)\,\Bigl(G^{AB}\,{\cal D}_A\Phi\star{\cal D}_B\Phi(\xi)+m^2\,
\Phi\star\Phi(\xi)\Bigr)
\label{scalaractiongen}\eeq
is gauge invariant if the scalar density $W(\xi)$ has the star-gauge
transformation property
\beq
\int\limits_{\reals^{2n}}d^{2n}\xi~
\Bigl(\delta_\alpha W(\xi)\,f(\xi)+W(\xi)~{\rm ad}_\alpha(f)(\xi)\Bigr)= 0
{}~~~~ \forall\alpha\in{\bf g} \ , f\in C^\infty(\real^{2n}) \ .
\label{conditionscalar}\eeq
Only a single function $W$ is required for the matter part of the action
because its Lorentz invariance properties will not become an issue in the
reduction. Note that the scalar fields decouple from the Yang-Mills fields in
the commutative limit where all $\Theta^{AB}$ vanish.

\newsection{Dimensional Reduction}

We will now describe a particular reduction of the generic noncommutative
Yang-Mills theory of the previous subsection. We will denote the local
coordinates of $\real^{2n}$ by $\xi=(x^\mu,y^a)$, where $\mu,a=1,\dots,n$,
and we break the Lorentz symmetry of $\real^{2n}$ to the direct product of the
Lorentz groups of $\real_x^{n}$ and $\real_y^{n}$. We will take the
noncommutativity parameters to be of the block form
\beq
\Theta^{AB}=\pmatrix{\theta^{\mu\nu}&\theta^{\mu b}\cr\theta^{a\nu}&
\theta^{ab}\cr}~~~~~~{\rm with}~~\theta^{\mu\nu}=\theta^{ab}=0 \ ,
\label{Thetadecomp}\eeq
and assume that $(\theta^{\mu b})$ is an invertible
$n\times n$ matrix. The flat metric of $\real^{2n}$ is taken to be
\beq
G^{AB}=\pmatrix{\eta^{\mu\nu}&\eta^{\mu b}\cr\eta^{a\nu}&
\eta^{ab}\cr}~~~~~~{\rm with}~~\eta^{\mu b}=\eta^{a\nu}=0 \ ,
\label{metricdecomp}\eeq
where $(\eta^{\mu\nu})=(\eta^{ab})={\rm diag}(1,-1,\ldots,-1)$. The vanishing
of the diagonal blocks $\theta^{\mu\nu}$ will be tantamount to the construction
of a quantum field theory on a {\it commutative} space $\real_x^{n}$. The $y^a$
can be interpreted as local coordinates on the cotangent bundle of
$\real_x^{n}$ (see the next section), so that the condition $\theta^{ab}=0$ is
tantamount to the commutativity of the corresponding ``momentum'' space. The
noncommutativity $\theta^{a\nu},\theta^{\mu b}$ between the coordinate and
``momentum'' variables will enable the construction of diffeomorphism
generators via star-commutators below. Having non-vanishing $\theta^{\mu\nu}$
would lead to some noncommutative field theory, but we shall not consider this
possibility here. In fact, in that case the noncommutative model only makes
sense in string theory~\cite{NCOS}, so that keeping $\theta^{\mu\nu}=0$ allows
us to define a quantum field theory in Minkowski signature without having to
worry about the problems of non-unitarity and non-covariance that plague
noncommutative field theories on non-Euclidean spacetimes. The bi-differential
operator (\ref{Deltadef}) which defines the star-product is then given by
\beq
\triangle=\frac 12\,\theta^{\mu a}\left(\overleftarrow{\partial}_\mu
\overrightarrow{\partial}_a-\overleftarrow{\partial}_a
\overrightarrow{\partial}_\mu\right) \ .
\label{Deltared}\eeq

Let us consider now the linear subspace ${\bf g}$ of smooth functions $\alpha$
on $\real^{2n}$ which are linear in the coordinates $y$,
\beq
\alpha(\xi)=\alpha_a(x)\,y^a \ .
\label{alphared}\eeq
Using (\ref{f1g1}) we then find that the star-commutator of any two elements
$\alpha,\beta\in {\bf g}$ is given by
\beq
[\alpha,\beta]_\star(\xi)&=&\Bigl([\alpha,\beta]_\star
\Bigr)_a(x)\,y^a \ , \nn\\
\Bigl([\alpha,\beta]_\star\Bigr)_a(x)&=&\theta^{\mu b}\,\Bigl[\beta_b(x)\,
\partial_\mu\alpha_a(x)-\alpha_b(x)\,\partial_\mu\beta_a(x)\Bigr] \ .
\label{vect}\eea
Thus $\bf g$ is a Lie algebra with respect to the star-commutator. If we now
define the invertible map
\bea
{\bf g}&\longrightarrow&\vect(\real_x^n)\nn\\
\alpha&\longmapsto&X_\alpha=-\theta^{\mu a}\,\alpha_a~\frac\partial
{\partial x^\mu}
\label{alphaVmap}\eea
onto the linear space of vector fields on $\real_x^n$, then (\ref{vect})
implies that it defines a representation of the Lie algebra $\bf g$,
\beq
[X_\alpha,X_\beta]=X_{[\alpha,\beta]_\star}~~~~\forall\alpha,\beta\in
{\bf g} \ .
\label{LieValpha}\eeq
This shows that, via the linear isomorphism (\ref{alphaVmap}), ${\bf g}$ can be
identified with the Lie algebra of connected diffeomorphisms of $\R_x^n$.

We now define a corresponding truncation of the space $\YM$ of Yang-Mills
fields on $\real^{2n}$ by
\beq
{\cal A}=\omega_{\mu a}(x)\,y^a~dx^\mu+C_a(x)~dy^a \ .
\label{gaugefieldred}\eeq
The reduction (\ref{gaugefieldred}) is the minimal consistent reduction which
is closed under the action of the reduced star-gauge group. It is
straightforward to compute the star-gauge transformations
(\ref{stargaugetransf}) of the component fields in (\ref{gaugefieldred}) using
the identities (\ref{f1g0}) and (\ref{f1g1}). One thereby checks that the
ansatz \Ref{gaugefieldred} is consistent, i.e. that the gauge transforms
$\delta_\alpha$ with gauge functions (\ref{alphared}) preserve the subspace of
$\YM$ of Yang-Mills fields of the form
(\ref{gaugefieldred}), and that the ``components'' of the gauge fields
transform as
\bea
\delta_\alpha\omega_{\mu a}&=&\partial_\mu\alpha_a+e\,\theta^{\nu b}
\,\left(\alpha_b\,\partial_\nu\omega_{\mu a}-\omega_{\mu b}\,\partial_\nu
\alpha_a\right) \ , \nn\\\delta_\alpha C_a&=&\alpha_a-
e\,\theta^{\mu b}\,\alpha_b\,\partial_\mu C_a \ , ~~ \alpha\in{\bf g} \ .
\label{gaugetransfred}\eea
The curvature components (\ref{fieldstrength}) of the gauge field
(\ref{gaugefieldred}) are likewise easily computed with the result
\bea
{\cal F}_{\mu\nu}(\xi)&=&\Omega_{\mu\nu a}(x)\,y^a \ , \nn\\
\Omega_{\mu\nu a}&=&
\partial_\mu\omega_{\nu a}-\partial_\nu\omega_{\mu a}+ e\,
\theta^{\lambda b}\left(\omega_{\nu b}\,\partial_\lambda\omega_{\mu a}-
\omega_{\mu b}\,\partial_\lambda\omega_{\nu a}\right) \ ,
\nn\\{\cal F}_{\mu a}&=&
\partial_\mu C_a-\omega_{\mu a}- e\,\theta^{\nu b}\,\omega_{\mu b}\,
\partial_\nu C_a \ , \nn\\{\cal F}_{ab}&=&0 \ .
\label{curvaturered}\eea

For the scalar fields, the consistent minimal truncation is to functions which
are independent of the $y$ coordinates,
\beq
\Phi(\xi) = \phi(x) \ .
\label{Phitrunc}\eeq
Using (\ref{f1g0}) the gauge transformation rule (\ref{Phigaugetransf}) then
implies
\beq
\delta_\alpha\phi=-\theta^{\mu a}\,\alpha_a\,\partial_\mu\phi \ .
\label{phigaugetransf}\eeq
Under the isomorphism ${\bf g}\cong\vect(\real_x^n)$ generated by
(\ref{alphaVmap}), we see that the gauge transform (\ref{phigaugetransf})
coincides with the standard transformation of a scalar field under
infinitesimal diffeomorphisms of $\real_x^n$, i.e. with the natural adjoint
action $\delta_\alpha\phi=X_\alpha(\phi)$ of $\vect(\real_x^n)$ on
$C^\infty(\real_x^n)$. The gauge covariant derivatives of the truncated
fields (\ref{Phitrunc}) are similarly easily computed to be
\bea
{\cal D}_\mu\Phi&=&\partial_\mu \phi - e\,
\theta^{\nu a}\,\omega_{\mu a}\,\partial_\nu \phi \ , \nn\\
{\cal D}_a\Phi&=&0 \ .
\label{DPhi}\eeq

It remains to compute the possible action functionals (\ref{NCYMactiongen}) and
(\ref{scalaractiongen}) corresponding to the above truncation. To arrive at a
gauge invariant action on $\R_x^n$, we make the ansatz
\bea
W^{\mu\mu'\nu\nu'}(\xi)&=&\eta^{\mu\mu'}\left[\eta^{\nu\nu'}\Bigl(w_\omega(x)
+\theta^{\lambda a}\,w^\omega_\lambda(x)\,\partial_a+\theta^{\lambda a}
\theta^{\lambda'b}\,w_{\lambda\lambda'}(x)\,\partial_a \partial_b\Bigr)
\right.\nn\\&&+\left.\theta^{\lambda a}\theta^{\lambda'b}\,
w^{\nu\nu'}_{\lambda\lambda'}(x)\,\partial_a \partial_b\right]~\delta^{(n)}(y)
\ , \nn\\W^{\mu\nu ab}(\xi)&=&\eta^{\mu\nu}\eta^{ab}\,w_C(x)~
\delta^{(n)}(y) \ , \nn\\W^{\mu\nu\nu'a}(\xi)&=&\eta^{\mu\nu}\,\theta^{\nu'a}
\,\theta^{\lambda b}\,w^M_\lambda(x)\,
\partial_b~\delta^{(n)}(y) \ , \nn\\W(\xi)&=&w_\phi(x)~\delta^{(n)}(y)
\label{Wxiansatz}\eea
for the weight functions. The functions $w$ in (\ref{Wxiansatz}) are smooth
functions in $C^\infty(\real_x^n)$, and the ansatz (\ref{Wxiansatz}) yields
well-defined action functionals over $\real_x^n$ provided that all component
fields live in an appropriate Schwartz subspace of $C^\infty(\real_x^n)$. The
choice (\ref{Wxiansatz}) of tensor densities represents a ``minimal''
dimensional reduction which is consistent with the reductions of the fields
above and which will naturally contain Einstein gravity in a particular limit.
There are of course many other choices for the functions $W^{AA'BB'}(\xi)$
which are possible, and these will lead to different types of diffeomorphism
invariant field theories. It is essentially here that there is the most freedom
involved. We have made the choice which will facilitate comparison to
previously known results in general relativity and in string theory. Due to the
structures of the spacetime metric (\ref{metricdecomp}), of the field strengths
(\ref{curvaturered}), and the symmetries (\ref{Wsyms}), the remaining
components of the tensorial weight functions in (\ref{Wxiansatz}) need not be
specified.

The derivative terms $\theta^{\lambda a}\,\partial_a$ in (\ref{Wxiansatz}) will
have the overall effect of transforming an $a$ index of $\real_y^n$ into a
$\lambda$ index of $\real_x^n$. The two choices of second order $y$-derivatives
in the first line of (\ref{Wxiansatz}) then correspond to the irreducible
decomposition of the reduced field strength tensors $\theta^{\lambda
a}\,\Omega_{\mu\nu a}$ under the action of the Lorentz group ${\rm SO}(1,n-1)$
of $\R_x^n$. These terms come from the rank two tensor ${\cal F}_{AB}$ of the
original noncommutative Yang-Mills theory which corresponds to the irreducible
antisymmetric representation $\Lambda^2(2n)$ of the Lorentz group ${\rm
SO}(1,2n-1)$. After dimensional reduction, it induces the rank $(1,2)$ tensor
$\theta^{\lambda a}\,\Omega_{\mu\nu a}$ which corresponds to the decomposable
representation
\beq
\Lambda^2(n)\otimes{\bf n}=\bar{\bf n}\oplus\bar{\bf n}\oplus
\Lambda_0^{1,2}(n) \ ,
\label{Lambda2nndecomp}\eeq
with $\bf n$ the defining and $\Lambda_0^{1,2}(n)$ the traceless, antisymmetric
$(1,2)$ representation of the reduced Lorentz group ${\rm SO}(1,n-1)$. In other
words, the restriction of the antisymmetric rank two representation of the
group ${\rm SO}(1,2n-1)$ to its ${\rm SO}(1,n-1)$ subgroup is reducible and
decomposes into irreducible representations according to
(\ref{Lambda2nndecomp}). The reduced Yang-Mills Lagrangian should be
constructed from Lorentz singlets built out of irreducible representations of
${\rm SO}(1,n-1)$. This requires the incorporation of the three ${\rm
SO}(1,n-1)$ singlets corresponding to the Clebsch-Gordan decomposition
(\ref{Lambda2nndecomp}). It is achieved by summing over the cyclic permutations
of the three indices of the reduced field strength tensor~\cite{TeleGR}, and
will be enforced by the given choice (\ref{Wxiansatz}).

The gauge transformation rules for the fields in (\ref{Wxiansatz}) can be
determined from the conditions (\ref{condition}) and (\ref{conditionscalar}).
Using these constraints it is straightforward to see that, for the types of
terms appearing in (\ref{Wxiansatz}), the index contractions specified there
are essentially unique, in that other choices are either forbidden by
star-gauge invariance or else they will produce the same local Lagrangian terms
in the end. In this sense, the ``minimal'' choice (\ref{Wxiansatz}) is unique
and star-gauge invariance forces very rigid constraints on the allowed tensor
weight functions. The restrictions (\ref{condition}) and
(\ref{conditionscalar}) are satisfied if the fields $w$ in (\ref{Wxiansatz})
transform as
\bea
\int\limits_{\reals^n}d^nx~\Bigl(\delta_\alpha w_\Xi(x)\,f(x,0)
+w_\Xi(x)~{\rm ad}_\alpha(f)(x,0)\Bigr)&=&0 \ ,
\label{conditionredw}\\ \int\limits_{\reals^n}d^nx~\theta^{\mu a}
\Bigl(\delta_\alpha w^\Xi_\mu(x)\,\partial_af(x,0)+
w^\Xi_\mu(x)\,\partial_a\,
{\rm ad}_\alpha(f)(x,0)\Bigr)&=&0 \ ,\label{conditionredwmu} \\
\int\limits_{\reals^n}d^nx~\theta^{\mu a}\theta^{\nu b}
\Bigl(\delta_\alpha w_{\mu\nu}(x)\,
\partial_a\partial_bf(x,0)+w_{\mu\nu}(x)\,\partial_a\partial_b\,
{\rm ad}_\alpha(f)(x,0)\Bigr)&=&0 \ , \label{conditionredwmunu}\\
\int\limits_{\reals^n}d^nx~\theta^{\mu a}\theta^{\nu b}
\Bigl(\delta_\alpha w_{\mu\nu}^{\lambda\lambda'}(x)\,
\partial_a\partial_bf(x,0)+w^{\lambda\lambda'}_{\mu\nu}(x)\,
\partial_a\partial_b\,{\rm ad}_\alpha(f)(x,0)\Bigr)&=&0 \ ,
\label{conditionred}\eea
for all smooth functions $f(x,y)$ which are compactly supported on $\real_x^n$
and quadratic in the $y^a$'s. The index $\Xi$ in (\ref{conditionredw}) denotes
the labels $\Xi=\omega,C,\phi$ while $\Xi=\omega,M$ in (\ref{conditionredwmu}).

We will solve (\ref{conditionredw})--(\ref{conditionred}) for the gauge
transformations of the functions $w$ appearing in (\ref{Wxiansatz}) by
demanding that these equations lead to local transforms of the fields $w$.
While the non-local integral transforms are required for the
distribution-valued densities $W$ on $\real^{2n}$, we will seek a dimensionally
reduced field theory in the following which possesses a local gauge symmetry.
For instance, setting $f(\xi)=f(x)$ independent of $y$ in
(\ref{conditionredw}), using (\ref{f1g0}), and integrating by parts over
$\real_x^n$ yields the local transforms
\beq
\delta_\alpha w_\Xi=-\partial_\mu\left(w_\Xi
\,\theta^{\mu a}\,\alpha_a\right) \ , ~~ \Xi=\omega,C,\phi \ .
\label{wgaugetransf}\eeq
Setting $f(\xi)=f_a(x)\,y^a$ linear in $y$ in (\ref{conditionredwmu}), using
(\ref{f1g1}), and integrating by parts over $\real_x^n$ yields
\beq
\delta_\alpha w^\Xi_\mu=-\partial_\nu\left(w^\Xi_\mu\,
\theta^{\nu a}\,\alpha_a\right)-\theta^{\nu a}\,w^\Xi_\nu\,
\partial_\mu\alpha_a \ , ~~ \Xi=\omega,M \ .
\label{wmugaugetransf}\eeq
Finally, setting $f(\xi)=f_{ab}(x)\,y^ay^b$ quadratic in $y$ in
(\ref{conditionredwmunu}) and (\ref{conditionred}), using (\ref{f2g1}) and
(\ref{wgaugetransf}), and integrating by parts over $\real_x^n$ gives
\bea
\delta_\alpha w_{\mu\nu}&=&-\partial_\lambda\left(w_{\mu\nu}\,
\theta^{\lambda a}\,\alpha_a\right)-
\theta^{\lambda a}\,w_{\lambda\nu}\,\partial_\mu\alpha_a-
\theta^{\lambda a}\,w_{\mu\lambda}\,\partial_\nu\alpha_a \ ,
\label{wmunugaugetransf}\\\delta_\alpha w^{\lambda\lambda'}_{\mu\nu}&=&
-\partial_\rho\left(w^{\lambda\lambda'}_{\mu\nu}\,
\theta^{\rho a}\,\alpha_a\right)-
\theta^{\rho a}\,w^{\lambda\lambda'}_{\rho\nu}\,\partial_\mu\alpha_a-
\theta^{\rho a}\,w^{\lambda\lambda'}_{\mu\rho}\,\partial_\nu\alpha_a \ .
\label{gammagaugetransf}\eea
It should be stressed that the transforms
(\ref{wgaugetransf})--(\ref{gammagaugetransf}) represent but a single solution
of the non-local constraint equations (\ref{condition}) and
(\ref{conditionscalar}). We have taken the solutions which will directly relate
local star-gauge invariance to general covariance in the dimensional reduction.

Using (\ref{f0g0}), (\ref{f1g1}), (\ref{curvaturered}) and (\ref{Wxiansatz}),
the noncommutative Yang-Mills action (\ref{NCYMactiongen}) can now be expressed
in terms of a local Lagrangian over the space $\real_x^n$ as
\beq
I_{\rm NCYM}^W=\int\limits_{\reals^n}d^nx~\Bigl(L_\omega+L_C+L_M\Bigr) \ ,
\label{INCYMWred}\eeq
where
\bea
L_{\omega}&=&\frac12\,\eta^{\mu\mu'}\,\theta^{\lambda a}\theta^{\lambda'b}\,
\left[2\,\eta^{\nu\nu'}\,w_{\lambda\lambda'}
\,\Omega_{\mu\nu a}\,\Omega_{\mu'\nu'b}+w_{\lambda\lambda'}^{\nu\nu'}\,
\Bigl(\Omega_{\mu\nu a}\,\Omega_{\mu'\nu'b}+
\Omega_{\mu\nu b}\,\Omega_{\mu'\nu'a}\Bigr)\right.\nn\\&&
-\frac12\,\eta^{\nu\nu'}\,w_\lambda^\omega\,\Bigl(\Omega_{\mu\nu a}\,
\partial_{\lambda'}\Omega_{\mu'\nu'b}-\Omega_{\mu'\nu'b}\,\partial_\lambda
\Omega_{\mu\nu a}\Bigr)\nn\\&&-\frac14\left.\eta^{\nu\nu'}\,w_\omega\,\Bigl(
\partial_{\lambda'}\Omega_{\mu\nu a}\Bigr)\Bigl(\partial_\lambda
\Omega_{\mu'\nu'b}\Bigr)\right] \ , \label{Lomega}\\
L_{C}&=&\frac12\,w_C\,\eta^{\mu\nu}\eta^{ab}\,{\cal F}_{\mu a}\,
{\cal F}_{\nu b} \ , \label{LC}\\L_M&=&\theta^{\nu'a}\theta^{\lambda b}\,
\eta^{\mu\nu}\,w_\lambda^M\,{\cal F}_{\mu a}\,\Omega_{\nu\nu'b} \ .
\label{LM}\eea
In a similar fashion the reduced scalar field action (\ref{scalaractiongen})
can be written as
\beq
I_{\rm B}=\int\limits_{\reals^n}d^nx~L_\phi \ ,
\label{IBred}\eeq
where
\beq
L_{\phi}=\frac12\,w_\phi\left(\eta^{\mu'\nu'}\,h_{\mu'}^\mu h_{\nu'}^{\nu}\,
(\partial_\mu \phi)(\partial_{\nu} \phi) + m^2\,\phi^2\right)
\label{Lphi}\eeq
and
\beq
h_\mu^\nu=\delta_\mu^\nu- e\,\theta^{\nu a}\,\omega_{\mu a} \ .
\label{hmunu}\eeq
In the following sections we will give geometrical interpretations of the field
theory (\ref{INCYMWred})--(\ref{hmunu}) and describe its relations to
gravitation.

\newsection{Induced Spacetime Geometry of Noncommutative Gauge Fields}

The remarkable property of the field theory of the previous section is that it
is diffeomorphism invariant. This follows from its construction and the
isomorphism (\ref{alphaVmap}), and is solely a consequence of the star-gauge
invariance of the original noncommutative Yang-Mills theory on $\real^{2n}$.
Precisely, it comes about from the representation (\ref{LieValpha}) of the Lie
algebra (\ref{deltastarcomm}) of star-gauge transformations in terms of vector
fields on flat infinite spacetime $\R_x^n$. This means that the various fields
induced in the previous section should be related in some natural way to the
geometry of spacetime. In this section we will show how this relationship
arises. We have already seen a hint of this diffeomorphism invariance in the
transformation law (\ref{phigaugetransf}) for the scalar fields, which we have
mainly introduced in the present context as source fields that probe the
induced spacetime geometry. The scalar field action (\ref{Lphi}) is in fact the
easiest place to start making these geometrical associations. This analysis
will clarify the way that the star-gauge symmetry of Yang-Mills theory on
noncommutative spacetime is related to the presence of gravitation.

The coordinates $y^a$ generate the algebra $C^\infty(\real_y^n)$ and obey the
star-commutation relations
\beq
\left[y^a\,,\,y^b\right]_\star=0 \ .
\label{yaybcomm}\eeq
Under a global coordinate translation $x^\mu\mapsto x^\mu+\epsilon^\mu$, the
scalar fields transform infinitesimally as
$\phi(x)\mapsto\phi(x)+\epsilon^\mu\,\partial_\mu\phi(x)$. Since
\beq
\partial_\mu\phi(x)=-\left(\theta^{-1}\right)_{a\mu}\,[y^a,\phi]_\star(x) \ ,
\label{innerderiv}\eeq
the derivative operator $\partial_\mu$ is an inner derivation of the algebra
$C^\infty(\real_x^n\times\real_y^n)$ and we may identify $y^a$ with the
holonomic derivative generators $-\theta^{\mu a}\,\partial_\mu$ of the
$n$-dimensional translation group ${\rm T}_n$ of $\real_x^n$. The standard,
flat space scalar field action $\int
d^nx~\frac12\,\eta^{\mu\nu}\,\partial_\mu\phi\,\partial_\nu\phi$ is invariant
under these global translations. Let us now promote the global ${\rm T}_n$
symmetry to a local gauge symmetry. This replaces global translations with
local translations $x^\mu\mapsto x^\mu+\epsilon^\mu(x)$ of the fiber
coordinates of the tangent bundle. It requires, in the usual way, the
replacement of the derivatives $\partial_\mu$ with the covariant derivatives
\beq
\nabla_\mu=\partial_\mu+e\,\omega_{\mu a}\,y^a \ ,
\label{nabladef}\eeq
where $\omega_{\mu a}$ are gauge fields corresponding to the gauging of the
translation group, i.e. to the replacement of $\real^n$ by the Lie algebra $\bf
g$ of local gauge transformations with gauge functions of the type
(\ref{alphared}). Using the identification (\ref{innerderiv}) it then follows
that the kinetic terms in the scalar field action will be constructed from
\beq
\nabla_\mu\phi=h_\mu^\nu\,\partial_\nu\phi \ ,
\label{nablaphi}\eeq
with $h_\mu^\nu$ given by (\ref{hmunu}). The covariance requirement
\beq
\delta_\alpha(\nabla_\mu\phi)=X_\alpha^\nu\,\partial_\nu(\nabla_\mu\phi)
\label{covnabla}\eeq
is equivalent to the gauge transformation law for the gauge fields
$\omega_{\mu a}$ in (\ref{gaugetransfred}).

The quantities (\ref{hmunu}) can thereby be identified with vierbein fields on
spacetime, and we see that the noncommutative gauge theory has the effect of
perturbing the trivial holonomic tetrad fields $\delta_\mu^\nu$ of flat space.
The noncommutative gauge fields become the non-trivial parts of the vierbein
fields and create curvature of spacetime. Note that in the present formalism
there is no real distinction between local spacetime and frame bundle indices,
because these are intertwined into the structure of the star-gauge group of the
underlying noncommutative gauge theory through the mixing of internal and
spacetime symmetries. In other words, the matrix $(\theta^{\mu a})$ determines
a linear isomorphism between the frame and tangent bundles of $\R_x^n$. It is
precisely this isomorphism that enables the present construction to go through.
We note also how naturally the identification (\ref{innerderiv}) of the
spacetime translational symmetry as an internal gauge symmetry arises from the
point of view of the original noncommutative Yang-Mills theory on $\real^{2n}$.
Using
(\ref{alphaVmap}) and (\ref{gaugetransfred}) we see that the identification of
(\ref{hmunu}) as a vierbein field is consistent with its gauge transform
\beq
\delta_\alpha h_\mu^\nu=X_\alpha^\lambda\,\partial_\lambda h_\mu^\nu-
h_\mu^\lambda\,\partial_\lambda X_\alpha^\nu
\label{hmunugaugetransf}\eeq
which coincides with the anticipated behaviour under infinitesimal
diffeomorphisms of~$\R_x^n$. The condition (\ref{hmunugaugetransf}) is
identical to the transformation law that one obtains from
(\ref{phigaugetransf}) and
the homogeneous transformation law (\ref{covnabla}) for the covariant
derivatives (\ref{nablaphi}). Note that $h^\nu_\mu$ behaves as a vector under
general coordinate transformations with respect to its upper index. As we will
discuss in the next section, it is a vector under local Lorentz transformations
with respect to its lower index.

We can now recognize the gauge transformation (\ref{wgaugetransf}) as the
infinitesimal diffeomorphism of a scalar density.\footnote{\baselineskip=12pt
For the function $w_\phi$ the condition (\ref{wgaugetransf}) may also be
naturally deduced from (\ref{phigaugetransf}) and by demanding that the mass
term of the Lagrangian (\ref{Lphi}) be invariant up to a total derivative under
infinitesimal diffeomorphisms.} Using (\ref{hmunugaugetransf}) this condition
can thereby be used to uniquely fix, up to a constant, the functions $w_\Xi$ in
terms of the noncommutative gauge fields, and we
have\footnote{\baselineskip=12pt Note that
$\det(h^\nu_\mu)=\sqrt{|\det(g_{\mu\nu})|}$ is the Jacobian of the frame bundle
transformation $\partial_\mu\mapsto\nabla_\mu$, where
$g_{\mu\nu}=\eta_{\mu'\nu'}\,h_\mu^{\mu'}h_\nu^{\nu'}$ is the Riemannian metric
induced by the vierbein fields.}
\beq
w_\Xi=\rho_\Xi~\det\left(h_\mu^\nu\right) \ , ~~ \Xi=\omega,C,\phi \ ,
\label{wXiNCYM}\eeq
where $\rho_\Xi$ are arbitrary constants. Similarly, the condition
(\ref{wmugaugetransf}) specifies that the functions $w_\mu^\Xi$ are vector
densities with respect to the connected diffeomorphism group of $\R_x^n$, and
from (\ref{hmunugaugetransf}) we may write
\beq
w_\mu^\Xi=\zeta_\Xi\, H_\mu^\nu\,\nabla_\nu\det\left(h_\lambda^{\lambda'}
\right)=\zeta_\Xi~\det\left(h_\lambda^{\lambda'}\right)~
 H_{\nu'}^\nu\,\partial_\mu h_\nu^{\nu'} \ , ~~ \Xi=\omega,M \ ,
\label{wmuXiNCYM}\eeq
where $\zeta_\Xi$ are arbitrary constants. Here $ H_\nu^\mu$ are the inverse
vierbein fields which are defined by the conditions
\beq
 H_\mu^\lambda\,h_\lambda^\nu=h_\mu^\lambda\, H_\lambda^\nu=\delta_\mu^\nu \ .
\label{invvierbeindef}\eeq
They are thereby determined explicitly in terms of the noncommutative gauge
fields as the perturbation series
\beq
H^\mu_\nu=\delta_\nu^\mu+e\,\theta^{\mu a}\,\omega_{\nu a}+
\sum_{k=2}^\infty e^k\,\theta^{\mu a_1}\theta^{\mu_1a_2}\cdots
\theta^{\mu_{k-1}a_k}\,\omega_{\mu_1a_1}\cdots
\omega_{\mu_{k-1}a_{k-1}}\omega_{\nu a_k} \ ,
\label{invvierexpl}\eeq
and they possess the infinitesimal gauge transformation property
\beq
\delta_\alpha H_\nu^\mu=-X_\alpha^\lambda\,\partial_\lambda
H_\nu^\mu- H_\lambda^\mu\,\partial_\nu X_\alpha^\lambda \ .
\label{deltaalphainvvier}\eeq
Finally, we come to the rank two tensor densities. From
(\ref{wmunugaugetransf}) we may identify
\beq
w_{\mu\nu}=\chi_0~\det\left(h_\lambda^{\lambda'}\right)~\eta_{\mu'\nu'}\,
 H_\mu^{\mu'}\, H_\nu^{\nu'} \ ,
\label{wmunuNCYM}\eeq
while from (\ref{gammagaugetransf}) we have
\beq
w_{\mu\nu}^{\lambda\lambda'}=\chi_{\bf n}~\det\left(h_{\mu'}^{\nu'}\right)~
 H_\mu^\lambda\, H_\nu^{\lambda'} \ ,
\label{wmunulambdaNCYM}\eeq
with $\chi_0$ and $\chi_{\bf n}$ arbitrary constants. As we shall see shortly,
the tensor density (\ref{wmunuNCYM}) is associated with the antisymmetric part
of the Clebsch-Gordan decomposition (\ref{Lambda2nndecomp}) while
(\ref{wmunulambdaNCYM}) is associated with the conjugate vector parts.

We see therefore that all fields of the previous section can be fixed in terms
of gauge fields of the dimensionally reduced noncommutative Yang-Mills theory.
All of the natural geometrical objects of spacetime are encoded into the
noncommutative gauge fields. Let us now consider the structure of the reduced
field strength tensor. From the form of the Lagrangian (\ref{Lomega}), and of
the weight functions (\ref{wXiNCYM}), (\ref{wmuXiNCYM}), (\ref{wmunuNCYM}) and
(\ref{wmunulambdaNCYM}), it follows that the natural objects to consider are
the contractions
\beq
T_{\mu\nu}^{~~\lambda}=-e\,\theta^{\lambda'a}\,H_{\lambda'}^\lambda\,
\Omega_{\mu\nu a}=H_{\lambda'}^\lambda\left(\nabla_\mu
h_\nu^{\lambda'}-\nabla_\nu h_\mu^{\lambda'}\right) \ ,
\label{Omegacontr}\eeq
with $\nabla_\mu=h_\mu^{\mu'}\,\partial_{\mu'}$. From (\ref{hmunugaugetransf})
and (\ref{deltaalphainvvier}) it follows that the curvatures (\ref{Omegacontr})
obey the homogeneous gauge transformation laws
\beq
\delta_\alpha T_{\mu\nu}^{~~\lambda}=X_\alpha^{\lambda'}\,
\partial_{\lambda'}T_{\mu\nu}^{~~\lambda} \ .
\label{Omegacontrgaugetransf}\eeq
{}From (\ref{Omegacontrgaugetransf}) one can check directly that each term in
the Lagrangian (\ref{Lomega}) is invariant up to a total derivative under
star-gauge transformations, as they should be by construction. From
(\ref{curvaturered}) and (\ref{hmunu}) it follows that the curvature
(\ref{Omegacontr}) naturally arises as the commutation coefficients in the
closure of the commutator of covariant derivatives to a Lie algebra with
respect to the given orthonormal basis of the frame bundle,
\beq
[\nabla_\mu,\nabla_\nu]=T_{\mu\nu}^{~~\lambda}\,\nabla_\lambda \ .
\label{curvLieT}\eeq
The operators $\nabla_\mu$ thereby define a non-holonomic basis of the tangent
bundle with non-holonomicity given by the noncommutative field strength tensor.
The change of basis $\nabla_\mu=h_\mu^\nu\,\partial_\nu$ between the coordinate
and non-coordinate frames is defined by the noncommutative gauge field.

The commutation relation (\ref{curvLieT}) identifies $T_{\mu\nu}^{~~\lambda}$,
or equivalently the noncommutative gauge field strengths $\Omega_{\mu\nu a}$,
as the torsion tensor fields of vacuum spacetime induced by the presence of a
gravitational field. The non-trivial tetrad field (\ref{hmunu}) induces a
teleparallel structure on spacetime through the Weitzenb\"ock connection
\beq
\Sigma^\lambda_{\mu\nu}= H_{\lambda'}^\lambda\,\nabla_\mu h_\nu^{\lambda'} \ .
\label{Weitzconn}\eeq
The connection (\ref{Weitzconn}) satisfies the absolute parallelism condition
\beq
{\sf D}_\mu(\Sigma)\,h_\nu^\lambda=
\nabla_\mu h_\nu^\lambda-\Sigma^{\lambda'}_{\mu\nu}\,h^\lambda_{\lambda'}=0 \ ,
\label{absparcond}\eeq
where ${\sf D}_\mu(\Sigma)$ is the Weitzenb\"ock covariant derivative. This
means that the vierbein fields define a mutually parallel system of local
vector fields in the tangent spaces of $\real_x^n$ with respect to the tangent
bundle geometry induced by $\Sigma^\lambda_{\mu\nu}$. The Weitzenb\"ock
connection has non-trivial torsion given by (\ref{Omegacontr}),
\beq
T_{\mu\nu}^{~~\lambda}=\Sigma^\lambda_{\mu\nu}-\Sigma^\lambda_{\nu\mu} \ ,
\label{torsionSigma}\eeq
but vanishing curvature,
\beq
R^{\mu'}_{~\nu'\mu\nu}(\Sigma)=\nabla_\mu\Sigma^{\mu'}_{\nu'\nu}-
\nabla_\nu\Sigma^{\mu'}_{\nu'\mu}+\Sigma^{\mu'}_{\lambda\mu}\,
\Sigma^{\lambda}_{\nu'\nu}-\Sigma^{\mu'}_{\lambda\nu}\,
\Sigma^{\lambda}_{\nu'\mu}=0 \ .
\label{Weitzcurv}\eeq

The teleparallel structure is related to a Riemannian structure on spacetime
through the identity
\beq
\Sigma_{\mu\nu}^\lambda=\Gamma_{\mu\nu}^\lambda+K_{\mu\nu}^{~~\lambda} \ ,
\label{WeitzLCrel}\eeq
where
\bea
\Gamma_{\mu\nu}^\lambda&=&\eta_{\mu'\nu'}\,\eta^{\sigma\sigma'}\,
H_\sigma^\lambda\, H_{\sigma'}^{\lambda'}\left(h^{\nu'}_\nu\,
\partial_\mu h^{\mu'}_{\lambda'}+h^{\mu'}_\mu\,\partial_\nu h^{\mu'}_{\lambda'}
-h^{\mu'}_\mu\,\partial_{\lambda'}h_\nu^{\nu'}-h^{\nu'}_\nu\,
\partial_{\lambda'}h_\mu^{\mu'}\right)\nn\\&&+\, H_{\lambda'}^\lambda\,
\partial_\mu h_\nu^{\lambda'}+ H_{\lambda'}^\lambda\,\partial_\nu
h_\mu^{\lambda'}
\label{LCconn}\eea
is the torsion-free Levi-Civita connection of the tangent bundle, and
\beq
K_{\mu\nu}^{~~\lambda}=\frac12\,\left(
\eta_{\mu\mu'}\,\eta^{\sigma\sigma'}\,H_\sigma^\lambda\,
H_{\sigma'}^{\lambda'}\,T_{\lambda'\nu}^{~~\mu'}+
\eta_{\nu\nu'}\,\eta^{\sigma\sigma'}\,H_\sigma^\lambda\,
H_{\sigma'}^{\lambda'}\,T_{\lambda'\mu}^{~~\nu'}-T_{\mu\nu}^{~~\lambda}\right)
\label{contorsion}\eeq
is the contorsion tensor. The torsion $T_{\mu\nu}^{~~\lambda}$ measures the
noncommutativity of displacements of points in the flat spacetime $\real_x^n$.
It is dual to the Riemann curvature tensor which measures the noncommutativity
of vector displacements in a curved spacetime. This follows from the identities
(\ref{Weitzcurv}) and (\ref{WeitzLCrel}) which yield the relationship
\bea
R^{\mu'}_{~\nu'\mu\nu}(\Gamma)&=&\partial_\mu\Gamma_{\nu'\nu}^{\mu'}-
\partial_\nu\Gamma^{\mu'}_{\nu'\mu}+\Gamma_{\lambda\mu}^{\mu'}\,
\Gamma_{\nu'\nu}^\lambda-\Gamma_{\lambda\nu}^{\mu'}\,
\Gamma_{\nu'\mu}^\lambda\nn\\&=&{\sf D}_\nu(\Gamma)\,K_{\nu'\mu}^{~~\mu'}
-{\sf D}_\mu(\Gamma)\,K_{\nu\nu'}^{~~\mu'}+K_{\nu'\mu}^{~~\lambda}\,
K_{\nu\lambda}^{~~\mu'}-K_{\nu'\nu}^{~~\lambda}\,K_{\mu\lambda}^{~~\mu'}
\label{RKrel}\eea
between the usual Riemann curvature tensor $R^{\mu'}_{~\nu'\mu\nu}(\Gamma)$ and
the torsion tensor. Here ${\sf D}_\nu(\Gamma)$ is the Riemannian covariant
derivative constructed from the Levi-Civita connection (\ref{LCconn}), whose
action on the contorsion tensor is given by
\beq
{\sf D}_\nu(\Gamma)\,K_{\nu'\mu}^{~~\mu'}=\partial_\nu K_{\nu'\mu}^{~~\mu'}+
\Gamma_{\lambda\nu}^{\mu'}\,K_{\nu'\mu}^{~~\lambda}+\Gamma_{\nu'\mu}^\lambda
\,K_{\lambda\nu}^{~~\mu'} \ .
\label{RcovK}\eeq

We see therefore that the dimensionally reduced noncommutative gauge theory of
the previous section gives a very natural model of a flat spacetime with a
given class of metrics carrying torsion, and with gauge field strengths
corresponding to the generic anholonomity of a given local orthonormal frame of
the tangent bundle of $\R_x^n$. It is precisely in this way that the
noncommutative gauge theory on flat spacetime can induce a model of curved
spacetime with torsion-free metric, i.e. it induces a teleparallel
Weitzenb\"ock geometry on $\R_x^n$ which is characterized by a
metric-compatible connection possessing vanishing curvature but non-vanishing
torsion and which serves as a measure of the intensity of the gravitational
field. The teleparallel structure naturally induces a Riemannian geometry on
spacetime, with curvature determined by the noncommutative field strength
tensor. As we have mentioned before, it is very natural that in a
noncommutative gauge theory, wherein global translations can be represented by
inner automorphisms of the algebra of functions on spacetime, the translation
group ${\rm T}_n$ be represented as an internal gauge symmetry group. In the
ensuing dimensional reduction it thereby becomes a genuine, local spacetime
symmetry of the field theory. The identification of the gauge field strengths
with torsion tensors is then also very natural, given the noncommutativity of
the spacetime coordinates and the fact that in noncommutative geometry the
star-product only yields a projective representation of the translation group
${\rm T}_n$ with cocycle determined by the noncommutativity parameters
$\Theta^{AB}$~\cite{LSZPrep}.

\newsection{Gravitation in Noncommutative Yang-Mills Theory}

We will now describe precisely how the diffeomorphism-invariant gauge field
theory (\ref{INCYMWred})--(\ref{LM}) is related to gravity in $n$ dimensions.
For this, we use the arbitrariness in the component weight functions to set the
higher derivative terms, i.e. the second and third lines of the Lagrangian
(\ref{Lomega}) to zero, $\rho_\omega=\zeta_\omega=0$. These terms represent
higher energy contributions to the field theory which admit a rather natural
string theoretical interpretation that we will describe in the next section.
Furthermore, we will see in section~7 that the Lagrangian (\ref{LC},\ref{LM})
for the auxilliary gauge fields $C_a(x)$ induces non-local interaction terms
for the gravitational gauge fields $\omega_{\mu a}$, and so also do not
contribute to the low-energy dynamics of the field theory (\ref{INCYMWred}). We
will therefore also set $\rho_C=\zeta_M=0$.

The low-energy dynamics of the dimensionally reduced noncommutative gauge
theory is thereby described by the Lagrangian
\beq
L_0=\frac1{2e^2}~\det\left(h_\sigma^{\sigma'}\right)~\eta^{\mu\mu'}
\left[2\chi_0\,\eta^{\nu\nu'}\,\eta_{\lambda\lambda'}\,T_{\mu\nu}^{~~\lambda}
\,T_{\mu'\nu'}^{~~\lambda'}+\chi_{\bf n}\left(T_{\mu\nu}^{~~\nu}\,
T_{\mu'\nu'}^{~~\nu'}+T_{\mu\nu}^{~~\nu'}\,T_{\mu'\nu'}^{~~\nu}\right)
\right] \ .
\label{L0}\eeq
The constant $\chi_0$ multiplies the torsion terms that arise from the
irreducible representation $\Lambda^{1,2}(n)$ in (\ref{Lambda2nndecomp}), while
the terms involving $\chi_{\bf n}$ come from the conjugate vector summands
$\bar{\bf n}$. The Lagrangian (\ref{L0}) belongs to the one-parameter family of
teleparallel Lagrangians (\ref{calLT},\ref{GRpareqns}) which describe
physically viable gravitational models, provided that the weight couplings obey
\beq
\chi_{\bf n}=-2\chi_0 \ .
\label{chitele}\eeq
The choice of constants (\ref{chitele}) as they appear in
(\ref{L0}) is quite natural from the point of view of the symmetries of the
Clebsch-Gordan decomposition (\ref{Lambda2nndecomp}). In this case, the
Lagrangian (\ref{L0}) represents a gravitational theory for macroscopic matter
which is observationally indistinguishable from ordinary general relativity.

This identification can be used to determine the Planck scale of the induced
gravitational model (\ref{L0}). For this, we note that with the choice of
weight functions (\ref{Wxiansatz}) the fields $\omega_{\mu a}$ have mass
dimension $\frac n2$ and the Yang-Mills coupling constant $e$ has mass
dimension $2-\frac n2$. In the gauge whereby the geometry is expanded around
flat spacetime, as in (\ref{hmunu}), the non-trivial parts of the vierbein
fields should assume the form $\kappa B^\nu_\mu$, where $\kappa$ is the Planck
scale and the translational gauge fields $B_\mu^\nu$ have mass dimension $\frac
n2-1$~\cite{Cho,Chogen}. To compare this with the perturbation $e\,\theta^{\nu
a}\,\omega_{\mu a}$ of the trivial tetrad field in (\ref{hmunu}), we introduce
the dimensionless noncommutativity parameters $\widehat{\theta}^{\mu
a}=\theta^{\mu a}/|\det(\theta^{\mu'a'})|^{1/n}$, which as discussed in the
previous section should be thought of, within the noncommutative geometry, as a
tensor mapping translation group valued quantities to quantities in the fiber
spaces of the
frame bundle. By comparing mass dimensions we see that we should then properly
identify
\beq
B_\mu^\nu=\left|\det\left(\theta^{\mu'a'}\right)\right|^{1/2n}~
\widehat{\theta}^{\nu a}\,\omega_{\mu a} \ .
\label{Bmunu}\eeq
Note that the Yang-Mills coupling constant itself cannot be used to compensate
dimensions, for instance in $n=4$ dimensions $e$ is dimensionless. Using
(\ref{Thetadecomp}), the Planck scale of $n$-dimensional spacetime is therefore
given in terms of $e$ and the noncommutativity scale as
\beq
\kappa=\sqrt{16\pi\,G_{\rm N}}=e~\left|{\rm Pfaff}
\left(\Theta^{AB}\right)\right|^{1/2n} \ .
\label{kappa}\eeq
Comparing (\ref{L0},\ref{chitele}) and (\ref{calLT}) then fixes the mass
dimension 2 weight constant $\chi_0$ to be
\beq
\chi_0=\left|{\rm Pfaff}\left(\Theta^{AB}\right)\right|^{-1/n} \ .
\label{chi0fix}\eeq
The induced gravitational constant (\ref{kappa}) vanishes in the
commutative limit and agrees with that found in~\cite{IIKKString} using the
supergravity dual of noncommutative Yang-Mills theory in four dimensions.

Let us now compare the low-energy field theory that we have obtained to
standard general relativity. By using the relation (\ref{WeitzLCrel}), the
Lagrangian
\beq
L_{\rm GR}=\frac{\chi_0}{e^2}~\det\left(h_\sigma^{\sigma'}\right)~
\eta^{\mu\mu'}\left[\frac14\,\eta^{\nu\nu'}\,\eta_{\lambda\lambda'}\,
T_{\mu\nu}^{~~\lambda}\,T_{\mu'\nu'}^{~~\lambda'}-T_{\mu\nu}^{~~\nu}\,
T_{\mu'\nu'}^{~~\nu'}+\frac12\,T_{\mu\nu}^{~~\nu'}\,T_{\mu'\nu'}^{~~\nu}\right]
\label{LG}\eeq
can be expressed in terms of the Levi-Civita connection
$\Gamma_{\mu\nu}^\lambda$ alone. By using (\ref{kappa},\ref{chi0fix}), along
with (\ref{contorsion}) and (\ref{RKrel}) to deduce the geometrical identity
\bea
R(\Gamma)&=&\eta^{\lambda\lambda'}\, H_\lambda^\nu\,
H_{\lambda'}^{\nu'}\,R^\mu_{~\nu'\mu\nu}(\Gamma)\nn\\
&=&\eta^{\mu\mu'}\left(T_{\mu\nu}^{~~\nu}\,
T_{\mu'\nu'}^{~~\nu'}-\frac12\,T_{\mu\nu}^{~~\nu'}\,T_{\mu'\nu'}^{~~\nu}-
\frac14\,\eta^{\nu\nu'}\,\eta_{\lambda\lambda'}\,T_{\mu\nu}^{~~\lambda}
\,T_{\mu'\nu'}^{~~\lambda'}\right.\nn\\&&+\biggl.\partial_\nu
K_{\mu\mu'}^{~~\nu}+K_{\mu\mu'}^{~~\nu}\, H_\sigma^{\sigma'}\,
\partial_\nu h_{\sigma'}^\sigma\biggr) \ ,
\label{Ridentity}\eea
the Lagrangian (\ref{LG}) can be rewritten, up to a total divergence, as the
standard Einstein-Hilbert Lagrangian
\beq
L_{\rm E}=-\frac1{16\pi\,G_{\rm N}}~\det\left(h_\lambda^{\lambda'}\right)~
R(\Gamma)
\label{LE}\eeq
in the first-order Palatini formalism. The Lagrangian (\ref{LG}) defines the
teleparallel formulation of general relativity, and it is completely equivalent
to Einstein gravity in the absence of spinning matter fields.

The main invariance property of the particular combination of torsion tensor
fields in (\ref{LG}) is its behaviour under a {\it local} change of frame
$\nabla_\mu$. This can be represented as a local Lorentz tranformation of the
vierbein fields
\bea
\delta^{\rm(L)}_\Lambda h_\mu^\nu(x)&=&\Lambda_\mu^{\mu'}(x)\,
h^\nu_{\mu'}(x) \ , \nn\\\delta^{\rm(L)}_\Lambda H_\nu^\mu(x)
&=&-\Lambda^\mu_{\mu'}(x)\, H_\nu^{\mu'}(x) \ ,
\label{deltaLh}\eeq
where $\Lambda_{\mu'}^\mu(x)$ are locally infinitesimal elements of ${\rm
SO}(1,n-1)$. Under (\ref{deltaLh}) the torsion tensor (\ref{Omegacontr})
transforms as
\beq
\delta^{\rm(L)}_\Lambda T_{\mu\nu}^{~~\lambda}=\partial_\mu\Lambda_\nu^\lambda
-\partial_\nu\Lambda_\mu^\lambda \ ,
\label{deltaLT}\eeq
from which it can be shown that the Lagrangian (\ref{LG}) changes by a total
derivative under a local change of frame~\cite{Cho}. The gauge field theory
defined by (\ref{LG}) is thereby independent of the choice of basis of the
tangent bundle used, and in particular of the decomposition (\ref{hmunu}) which
selects a gauge choice for the vierbein fields corresponding to a background
perturbation of flat spacetime. The field equations derived from (\ref{LG})
will then uniquely determine the spacetime geometry and hence the orthonormal
teleparallel frame up to a global Lorentz transformation. In fact, the
Einstein-Hilbert Lagrangian defines the unique teleparallel gravitational
theory which possesses this local Lorentz invariance~\cite{Cho}.

The Lagrangian (\ref{L0}), on the other hand, is only invariant under {\it
global} Lorentz transformations. This owes to the fact that the original
noncommutative gauge theory on $\real^{2n}$ is only invariant under a flat
space Lorentz group, and in the dimensional reduction only the translational
subgroup of the full Poincar\'e group is gauged to a local symmetry. The
reduced gauge theory is thereby a dynamical theory of the prefered orthonormal
teleparallel frames in which the connection coefficients vanish and the torsion
tensor has the simple form (\ref{Omegacontr}). The frame $\nabla_\mu$ is then
specified only modulo some local Lorentz transformation and the parallelism is
not uniquely determined. The gravitational model is therefore ambiguous because
there is a whole gauge equivalence class of geometries representing the same
physics. Nevertheless, with the choice of parameters
(\ref{chitele},\ref{chi0fix}), the Lagrangian (\ref{L0}) lies in the
one-parameter family of teleparallel theories (\ref{calLT},\ref{GRpareqns})
which pass all observational and theoretical tests of Einstein gravity. We use
this criterion to fix the arbitrary constants of the gravitational model
(\ref{L0}), whose presence effectively encodes the long distance effects of the
internal space $\real_y^n$.

\newsection{D-Branes and Volume Preserving Diffeomorphisms}

To understand what the higher derivative terms in the Lagrangian (\ref{Lomega})
represent, we return to the standard action (\ref{NCYMstandard}) for
noncommutative Yang-Mills theory on $\real^{2n}$. This is the action that is
induced on a flat D$(2n-1)$-brane in flat space and in the presence of a
constant background $B$-field. We can now examine the ``naive'' dimensional
reduction of this action to an $n$-dimensional submanifold
$\real_x^n\subset\real^{2n}$. Such a submanifold could correspond, for example,
to the embedding of a flat D$(n-1)$-brane inside the D$(2n-1)$-brane with a
transverse $B$-field, which realizes the D$(n-1)$-brane as a noncommutative
soliton in the worldvolume of the D$(2n-1)$-brane~\cite{MPT}. The reduced
action is then given by
\beq
I_{\rm NCYM}^{\rm red}=\frac{{\rm vol}_y}2~\int\limits_{\reals^n}
d^nx~G^{AA'}G^{BB'}\,{\cal F}_{AB} \star {\cal F}_{A'B'}(x,0) \ ,
\label{NCYMstandardred}\eeq
where ${\rm vol}_y$ is the volume of the transverse space. Note that in this
reduction the mass dimension of the Yang-Mills coupling constant $e$ is $2-n$.

Within the general formalism of section~3, it follows that we should choose all
$w_\Xi={\rm vol}_y$, while
$w_\mu^\Xi=w_{\mu\nu}=w_{\mu\nu}^{\lambda\lambda'}=0$. The field theory
(\ref{INCYMWred}) is then given by the local Lagrangian
\beq
L_D=-\frac{{\rm vol}_y}{2e^2}\,\eta^{\mu\mu'}\,\left[\eta^{\nu\nu'}\,
\partial_{\lambda'}\Bigl(h^\lambda_\sigma\,T_{\mu\nu}^{~~\sigma}\Bigr)\,
\partial_\lambda\Bigl(h^{\lambda'}_{\sigma'}\,T_{\mu'\nu'}^{~~\sigma'}\Bigr)
-e^2\,\eta^{aa'}\,{\cal F}_{\mu a}\,{\cal F}_{\mu'a'}\right] \ .
\label{LD}\eeq
However, in order for (\ref{LD}) to define a diffeomorphism invariant field
theory, we still need to satisfy the star-gauge invariance conditions
(\ref{condition}). While the transforms are trivially satisfied of course for
the vanishing $w$'s, the constraint (\ref{wgaugetransf}) for constant $w_\Xi$
imposes the restriction
\beq
\partial_\mu X_\alpha^\mu=0
\label{volpreseqn}\eeq
on the types of diffeomorphisms which can be used for star-gauge
transformations. This means that the map (\ref{alphaVmap}) is not surjective
and its image consists of only volume preserving diffeomorphisms. This is
expected from the fact that the component functions of the weight densities are
constant, so that the only diffeomorphism invariance that one can obtain in
this case are the coordinate transformations that leave the flat volume element
of $\real_x^n$ invariant, i.e. those which are isometries of the flat Minkowski
metric $\eta_{\mu\nu}$ and thereby infinitesimally satisfy (\ref{volpreseqn}).
Thus, in the D-brane interpretation of the dimensional reduction presented in
this paper, star-gauge invariance acts to partially gauge fix the
diffeomorphism group of spacetime. One arrives at not a full theory of gravity,
but rather one which is only invariant under the subgroup consisting of volume
preserving diffeomorphisms. This subgroup arises as the residual symmetry of
the field theory that remains after the gauge-fixing.

Generally, volume preserving diffeomorphisms constitute the symmetry group
which reflects the spacetime noncommutativity that arises in D-brane
models~\cite{Matsuo}. For instance, they arise as the dynamical degree of
freedom in matrix models~\cite{Matrix} which comes from the discretization of
the residual gauge symmetry of the 11-dimensional supermembrane~\cite{dWHN}.
They also appear as the residual symmetry after light-cone gauge fixing in
$p$-brane theories~\cite{pbrane}, and they naturally constitute the Lie algebra
of star-gauge transformations in noncommutative Yang-Mills theory on flat
spacetime~\cite{LSZPrep}. Here we have tied them in with the dynamics of
D-branes through the effective, higher-derivative gravitational theories
(\ref{LD}) that are induced in the dimensional reduction. Another way to see
that general covariance in the usual noncommutative gauge theories is only
consistent with volume preserving symmetries is by noting the infinitesimal
coordinate transformation $\delta_\alpha x^\mu=X_\alpha^\mu(x)$ implies that
the noncommutativity parameters $\theta^{\mu a}=[x^\mu,y^a]_\star$ must
transform under gauge transformations as
\beq
\delta_\alpha\theta^{\mu a}=[X_\alpha^\mu,y^a]_\star=\theta^{\nu a}\,
\partial_\nu X_\alpha^\mu \ .
\label{deltaalphatheta}\eeq
Requiring that the noncommutative gauge symmetries preserve the supergravity
background on the D-branes sets (\ref{deltaalphatheta}) to zero, which also
leads to the isometry constraint above. This constraint further ensures that
the tensor $\theta^{\mu a}$ defines a global isomorphism between the frame and
tangent bundles, as is required in the construction of this paper.

The higher derivative terms in the action (\ref{Lomega}) can thereby be thought
of as ``stringy'' corrections to the teleparallel gravity theory. It is
tempting to speculate that they are related to the higher-curvature couplings
that arise in effective supergravity actions. It is a curious fact that in this
interpretation one does not arrive at an Einstein-like theory of gravity on the
D-brane. This induced brane gravity deserves to be better understood, and most
notably how the analysis of this section and the previous one relates to the
large $N$ supergravity results which demonstrate the existence of conventional
gravitation in noncommutative gauge theory~\cite{IIKKString}. The dimensional
reductions of section~5 could indeed be related to the way that the Newtonian
gravitational potential arises from a Randall-Sundrum type localization on
anti-de~Sitter space. Indeed, it would be interesting to understand whether or
not the generalized class of noncommutative gauge theories
(\ref{NCYMactiongen}) arises as an effective field theory of strings in some
limit, or if the dimensional reductions follow from some sort of dynamical
symmetry breaking mechanism in the noncommutative quantum field theory. This
would presumably fix all
free parameters of the induced gravitational theory~(\ref{INCYMWred}).

\newsection{Role of the Auxilliary Fields}

The most important ingredient missing from the induced gravity model of
section~5 is local Lorentz invariance. This somewhat undesirable feature owes
to the indistinguishability within the present formalism between spacetime and
frame indices. It is in fact quite natural from the point of view of the
original noncommutative gauge theory, whereby the star-gauge symmetry allows
the gauging of the translation group but is independent of the invariance of
the field theory under global ${\rm SO}(1,2n-1)$ transformations. We may
expect, however, that local Lorentz symmetry is restored in some complicated
dynamical way in the reduced noncommutative gauge theory, such that the
effective gauge theory contains general relativity. This problem is addressed
in~\cite{AIKO} in the context of reduced models. In this section we will
briefly describe some potential steps in this direction.

The natural place to look for the extra terms required to make the Lagrangian
(\ref{L0}) invariant under local frame rotations is in the terms involving the
auxilliary ``internal'' gauge fields $C_a(x)$, whose role in the induced
gravitational theory has thus far been ignored. They represent the components
of the noncommutative gauge field in the internal directions, along which lies
the coordinate basis $y^a$ defining the generators of the translation group
${\rm T}_n$ that is used in the gauging prescription. They thereby represent
natural candidates to induce the necessary terms that instate the frame basis
independence of the diffeomorphism invariant field theory of section~5. Note
that these fields cannot be set to zero because of their gauge transformation
law in (\ref{gaugetransfred}). They therefore constitute an intrinsic dynamical
ingredient of the induced gravity model.

The variation of the Lagrangian $L_C+L_M$ given by (\ref{curvaturered}),
(\ref{LC}), (\ref{LM}), (\ref{wXiNCYM}) and (\ref{wmuXiNCYM}) with respect to
the auxilliary gauge fields $C_a(x)$ yields the field equation
\beq
\eta^{ab}\,\Box C_b(x)= J^a(x) \ ,
\label{fieldeq}\eeq
where we have introduced the second order linear differential operator
\beq
\Box=\rho_C\,\eta^{\mu\nu}\left[
\left(\partial_\lambda h_\mu^\lambda\right)\nabla_\nu+\left(\nabla_\mu
h_\nu^\lambda\right)H_\lambda^{\nu'}\,\nabla_{\nu'}+ H_\lambda^{\mu'}
\left(\nabla_\mu h_{\mu'}^\lambda\right)\nabla_\nu+h_\nu^\lambda\,
\nabla_\mu\,\partial_\lambda\right]
\label{Box}\eeq
and the fields
\bea
 J^a&=&\eta^{\mu\nu}\left\{\rho_C\,
\eta^{ab}\left[ H_\lambda^{\mu'}\left(\nabla_\nu h_{\mu'}^\lambda
\right)\,\omega_{\mu b}+\omega_{\mu b}\,\partial_\lambda h^\lambda_\nu+
\nabla_\mu\omega_{\nu a}\right]\nn\right.\\&&+\,\frac{\zeta_M}e\,
\theta^{\nu'a}\,\left[ H_{\lambda'}^{\mu'}\,
T_{\nu\nu'}^{~~\lambda}\left(\nabla_\lambda h^{\lambda'}_{\mu'}\right)
\left(\partial_{\nu''}h^{\nu''}_\mu\right)+ H^{\mu'}_{\lambda'}\,
 H^{\lambda''}_{\nu''}\,T_{\nu\nu'}^{~~\lambda}
\left(\nabla_\lambda h_{\mu'}^{\lambda'}\right)\left(\nabla_\mu
h_{\lambda''}^{\nu''}\right)\right.\nn\\&&+\left.\left.
H^{\mu'}_{\lambda'}\,\nabla_\mu\left(h_{\lambda''}^\lambda\,
T_{\nu\nu'}^{~~\lambda''}\right)\left(\partial_\lambda h_{\mu'}^{\lambda'}
\right)+h^\lambda_{\lambda''}\,T_{\nu\nu'}^{~~\lambda''}\,\nabla_\mu
\left( H_{\lambda'}^{\mu'}\,\partial_\lambda h_{\mu'}^{\lambda'}\right)
\right]\right\} \ .
\label{Sigmaa}\eea
Substituting the solution of (\ref{fieldeq}) for the fields $C_a(x)$ into the
Lagrangian $L_C+L_M$ thereby yields the non-local effective Lagrangian
\bea
L_{\rm eff}&=&\det\left(h_\sigma^{\sigma'}\right)~\left[
-\frac12\,\eta_{ab}\, J^a\,\frac1\Box J^b+\frac{\rho_C}2\,
\eta^{\mu\nu}\,\eta^{ab}\,\omega_{\mu a}\,\omega_{\nu b}
\right.\nn\\&&+\left.\frac{\zeta_M}e\,
\theta^{\nu'a}\,\eta^{\mu\nu}\, H_{\mu'}^{\lambda'}\,
\left(\nabla_\lambda h_{\lambda'}^{\mu'}\right)\,
\omega_{\mu a}\,T_{\nu\nu'}^{~~\lambda}\right] \ .
\label{Leff}\eea
Note that in the case $\rho_C=0$, the auxilliary fields are Lagrange
multipliers which enforce a geometric constraint given by setting the fields
(\ref{Sigmaa}) identically equal to zero.

By performing a gradient expansion of the operator $\Box^{-1}$, we can now
study the derivative expansion of the effective Lagrangian (\ref{Leff}).
Higherderivative terms can be attributed to stringy corrections, as they were
in the
previous section. The leading order terms may then lead to the appropriate
additions of terms to the Lagrangian (\ref{L0}) which makes it invariant under
local Lorentz transformations. However, generically the Lagrangian (\ref{Leff})
will also contain infinitely many higher derivative terms and so a minimal,
low-energy model is not strictly speaking attainable with this reasoning. In
fact, one can simply set the constants $\rho_C=\zeta_M=0$ and completely ignore
the non-local contributions from the auxilliary fields. Their inclusion
represents the possibility of obtaining a gravitational field theory which is
completely equivalent to general relativity, at least on a macroscopic scale.
This possibility deserves further investigation.

\newsection{Coupling Gauge Theory and Gravity}

For gauge functions of the form (\ref{alphared}), we have used a principle of
``minimal consistent reduction'' to fix the fields that arise in the induced
gravitational theory. This led to the choice (\ref{gaugefieldred}) involving
the gauge fields $\omega_{\mu a}$ which induced the non-trivial part of the
tetrad fields of the induced spacetime geometry, and the auxilliary fields
$C_a$. It is possible to consider more general Lie algebras $\bf g$ other than
the minimal one consisting of the gauge functions (\ref{alphared}). For
instance, it is possible to define $\alpha(\xi)$ with a piece $\alpha^{(0)}(x)$
which is independent of the $y$ coordinates,
\beq
\alpha(\xi)=\alpha^{(0)}(x)+\alpha_a(x)\,y^a \ .
\label{alphagen}\eeq
Again $\bf g$ is a Lie algebra with respect to the star-commutator, with
``component'' functions (\ref{vect}) and
\bea
\Bigl([\alpha,\beta]_\star\Bigr)^{(0)}(x)
&=&\theta^{\mu a}\left[\beta_a(x)\,\partial_\mu\alpha^{(0)}(x)
-\alpha_a(x)\,\partial_\mu\beta^{(0)}(x)\right]\nn\\&=&
X_\alpha^\mu(x)\,\partial_\mu\beta^{(0)}(x)-X_\beta^\mu(x)\,
\partial_\mu\alpha^{(0)}(x)
\label{alphabetacomm0}\eea
for $\alpha,\beta\in{\bf g}$. The smallest truncation of the space $\YM$ is now
defined by Yang-Mills fields of the form
\beq
{\cal A}=\Bigl(A_\mu(x)+\omega_{\mu a}(x)\,y^a\Bigr)~dx^\mu+C_a(x)~dy^a \ ,
\label{calArednew}\eeq
and the star-gauge transformation rules (\ref{gaugetransfred}) are supplemented
with
\bea
\delta_\alpha A_\mu&=&\partial_\mu\alpha^{(0)}+e\,\theta^{\lambda a}\,
\left(\alpha_a\,\partial_\lambda A_\mu-
\omega_{\mu a}\,\partial_\lambda\alpha^{(0)}\right)\nn\\&=&\nabla_\mu
\alpha^{(0)}-e\,X_\alpha^\nu\,\partial_\nu A_\mu \ .
\label{deltaalphaamu}\eeq
The noncommutative field strength tensor is then modified as
\bea
{\cal F}_{\mu\nu}(\xi)&=&F_{\mu\nu}(x)+e\,\theta^{\lambda a}\,
\Bigl(\omega_{\nu a}(x)\,\partial_\lambda A_\mu(x)-\omega_{\mu a}(x)\,
\partial_\lambda A_\nu(x)\Bigr)+\Omega_{\mu\nu a}(x)\,y^a\nn\\&=&
\nabla_\mu A_\nu(x)-\nabla_\nu A_\mu(x)+\Omega_{\mu\nu a}(x)\,y^a \ ,
\label{calFmod}\eea
where
\beq
F_{\mu\nu}=\partial_\mu A_\nu-\partial_\nu A_\mu
\label{fmununew}\eeq
and the remaining components of ${\cal F}_{AB}$ are as in (\ref{curvaturered}).

It follows that the choice (\ref{alphagen}) of gauge functions induces a model
of ordinary Maxwell electrodynamics for the photon field $A_\mu(x)$ on
$\real_x^n$ coupled to gravity. Note that the star-gauge invariance of the
original Yang-Mills theory mixes up the ${\rm U}(1)$ internal symmetries with
the spacetime symmetries, as is evident in the expressions
(\ref{alphabetacomm0}), (\ref{deltaalphaamu}) and (\ref{calFmod}). In
particular, from (\ref{deltaalphaamu}) we see that the photon field $A_\mu$
transforms covariantly under general coordinate transformations, while it is a
vector under the local Lorentz group,
\beq
\delta_\Lambda^{\rm (L)}A_\mu(x)=\Lambda^\nu_\mu(x)\,A_\nu(x) \ .
\label{deltaLAmu}\eeq
In this way one obtains a unified gauge theory which couples the gravitational
theory that was studied in earlier sections to electrodynamics. Notice also
that the photon field $A_\mu(x)$ does not couple to the scalar field $\phi(x)$,
consistent with the fact that the scalar bosons are taken to be neutral under
the extra abelian gauge symmetry. This generalization evidently also goes
through if one starts from a noncommutative Yang-Mills theory with some
non-abelian ${\rm U}(N)$ gauge group. Then one obtains a sort of non-abelian
model of gravity coupled to ordinary Yang-Mills theory. However, the star-gauge
group of the simpler noncommutative electrodynamics contains all possible
non-abelian unitary gauge groups in a very precise way~\cite{LSZPrep,GN2}. It
would be interesting then to extract the gravity-coupled Yang-Mills theory
directly from a dimensionally reduced gauge theory of the form
(\ref{NCYMactiongen}). There is therefore a wealth of gravitational theories
that can be induced from noncommutative gauge theory, which in itself also
seems to serve as the basis for a unified field theory of the fundamental
forces. In all instances the type of theory that one obtains is dictated by the
choice of reduced star-gauge group, i.e. the Lie algebra $\bf g$, as well as
the choice of weight functions $W$ for the dimensional
reduction. This illustrates the richness of the constraints of star-gauge
invariance in noncommutative Yang-Mills theory.

\newsection{Conclusions}

In this paper we have described a particular class of dimensional reductions of
noncommutative electrodynamics which induce dynamical models of spacetime
geometry involving six free parameters. Two of these parameters can be fixed by
requiring that the leading, low-energy dynamics of the model be empirically
equivalent to general relativity. The higher-order derivative corrections can
be attributed to stringy corrections and non-local effects due to
noncommutativity. The low-energy dynamics can be consistently decoupled from
the high-energy modes by an appropriate choice of parameters. These results
show that a certain class of teleparallel gravity theories have a very natural
origin in a noncommutative gauge theory whereby diffeomorphism invariance is
solely a consequence of the star-gauge invariance of the Yang-Mills theory, in
the same spirit as the usual gauge theories based on the translation group of
flat space. Alternatively, the present construction sheds light on the manner
in which noncommutative gauge theories on flat spacetime contain gravitation.
We have also described how Yang-Mills theory on a noncommutative space
naturally contains a gravitational coupling of ordinary gauge theories to the
geometrical model studied in most of this paper. A real advantage of this point
of view of inducing gravity from noncommutative gauge theory is that in the
latter theory it is straightforward to construct gauge-invariant observables.
These are constructed in terms of the open and closed Wilson line operators,
which are non-local in character. It would be interesting to understand these
observables from the point of view of the induced gravitational theory.

It should be stressed that we have only presented a very simple model of
dimensional reduction. More general reductions are possible and will induce
different geometrical models. The present technique can be regarded as a
systematic way to induce theories of gravitation starting only from the single,
elementary principle of star-gauge invariance of noncommutative Yang-Mills
theory. One extension would be to include a gauging of the full Poincar\'e
group of spacetime. This would cure the problem of local Lorentz invariance and
potentially yield a theory of gravitation which is completely equivalent to
general relativity. It should be possible to find such an extended
noncommutative gauge theory whose dimensional reduction yields the appropriate
model with manifest local Lorentz symmetry. After an appropriate gauge fixing,
this model should then reduce to the theory analysed in this paper.

\setcounter{section}{0}

\subsection*{Acknowledgments}

E.L. thanks I. Bengtsson and J. Mickelsson for helpful discussions. This work
was supported in part by the Swedish Natural Science Research Council (NFR).
Financial support of the ``G\"oran Gustafsson Stiftelse'' is also
gratefully acknowledged. The work of R.J.S. was supported in part by an
Advanced Fellowship from the Particle Physics and Astronomy Research
Council~(U.K.).

\appendix{Reduced Star Product Identities}

In this Appendix we collect, for convenience, a few formulas which were used
to derive the equations given in the text. Let $\Theta^{AB}$ be as in
(\ref{Thetadecomp}). We denote by $f^{(k)}\in C^\infty(\R_x^{n}\times
\R_y^{n})$, $k\geq0$, a function which is of degree $k$ in the $y$ coordinates,
\beq
f^{(k)}(\xi)=f_{a_1\cdots a_k}(x)\,y^{a_1}\cdots y^{a_k} \ ,
\label{funred}\eeq
where $f_{a_1\cdots a_k}(x)$ is completely symmetric in its indices,
and $f^{(0)}(\xi)=f(x)$ is independent of $y$.

We then have the following reduced star-product identities:
\bea
f^{(0)}\star g^{(0)}(\xi)&=&f(x)g(x) \ ; \label{f0g0}\\ ~~~&&~~~ \nn\\
f^{(1)}\star g^{(0)}(\xi)&=&g(x)f_a(x)\,y^a-
\frac12\,\theta^{\mu a}\,f_a(x)\,\partial_\mu g(x)  \ , \nn \\
g^{(0)}\star f^{(1)}(\xi)&=&g(x)f_a(x)\,y^a+
\frac12\,\theta^{\mu a}\,f_a(x)\,\partial_\mu g(x) \ ; \label{f1g0}
\\ ~~~&&~~~ \nn\\
f^{(1)}\star g^{(1)}(\xi)&=&f_a(x) g_b(x) \,y^a y^b + \frac12\,\theta^{\mu a}
\left[\Bigl(\partial_\mu f_b(x)\Bigr) g_a(x) \nn \right.\\&&
-\left.f_a(x)\Bigl(\partial_\mu g_b(x)\Bigr)\right]y^b -
\frac14\,\theta^{\mu a}\theta^{\nu b}\Bigl(\partial_\nu f_a(x)\Bigr)
\Bigl(\partial_\mu g_b(x)\Bigr) \ ; \label{f1g1}\\ ~~~&&~~~ \nn\\
\left[g^{(1)}\,,\,f^{(1)}\star f^{(1)}\right]_\star(\xi)&=&2\theta^{\mu a}
\Bigl(\partial_\mu g_b(x)\Bigr)f_a(x)f_c(x)\,y^by^c+\Bigl(\theta^{\lambda a}\,
\partial_\lambda g_a(x)\Bigr)\biggl\{f_b(x)f_c(x)\,y^by^c\biggr.\nn\\& &
+\left.\frac{\theta^{\mu b}\theta^{\nu c}}4\,\left[\partial_\mu\partial_\nu
\Bigl(f_b(x)f_c(x)\Bigr)+\Bigl(\partial_\nu f_b(x)\Bigr)\Bigl(
\partial_\mu f_c(x)\Bigr)\right]\right\}\nn\\& &+\,\frac{\theta^{\lambda a}}4
\,\partial_\lambda\left[2 g_a(x)f_b(x)f_c(x)\,y^by^c\right.\nn\\&&-\,
\theta^{\mu b}\theta^{\nu c}\, g_a(x)\Bigl(\partial_\nu f_b(x)\Bigr)\Bigl(
\partial_\mu f_c(x)\Bigr)\nn\\& &-\,\theta^{\mu b}\theta^{\nu c}\,
\Bigl(\partial_\nu g_b(x)\Bigr)\partial_\mu\Bigl(f_a(x)f_c(x)\Bigr)\nn\\&&+\,
\theta^{\mu b}\theta^{\nu c}\, g_b(x)\,\partial_\mu\partial_\nu
\Bigl(f_a(x)f_c(x)\Bigr)\nn\\& &-\left.\theta^{\mu b}\theta^{\nu c}\,
 g_a(x)\,\partial_\mu\partial_\nu\Bigl(f_b(x)f_c(x)\Bigr)\right] \ ;
\label{f1f1g1}\\ ~~~&&~~~ \nn\\
\left[g^{(1)}\,,\,f^{(2)}\right]_\star(\xi)&=&\theta^{\mu a}\Bigl(
2f_{ac}(x)\,\partial_\mu g_b(x)-g_a(x)\,\partial_\mu f_{bc}(x)
\Bigr)\,y^by^c\nn\\&&-\,\theta^{\mu a}\theta^{\nu b}\theta^{\lambda c}
\Bigl(\partial_\mu\partial_\nu g_c(x)\Bigr)\Bigl(\partial_\lambda f_{ab}(x)
\Bigr) \ .
\label{f2g1}\eea

\end{document}